\documentclass[final,5p,times,twocolumn]{elsarticle}

\usepackage{graphicx}
\usepackage{amssymb}
\journal{Nuclear Instruments and Methods A}

\begin{document}
%\linenumbers. 
\begin{frontmatter}

%% Title, authors and addresses

%% use the tnoteref command within \title for footnotes;
%% use the tnotetext command for the associated footnote;
%% use the fnref command within \author or \address for footnotes;
%% use the fntext command for the associated footnote;
%% use the corref command within \author for corresponding author footnotes;
%% use the cortext command for the associated footnote;
%% use the ead command for the email address,
%% and the form \ead[url] for the home page:
%%
%% \title{Title\tnoteref{label1}}
%% \tnotetext[label1]{}
%% \author{Name\corref{cor1}\fnref{label2}}
%% \ead{email address}
%% \ead[url]{home page}
%% \fntext[label2]{}
%% \cortext[cor1]{}
%% \address{Address\fnref{label3}}
%% \fntext[label3]{}

\title{Development of a novel scintillation-trigger detector for the MTV experiment \\using aluminum-metallized film tapes}

%% use optional labels to link authors explicitly to addresses:
%% \author[label1,label2]{<author name>}
%% \address[label1]{<address>}
%% \address[label2]{<address>}

\author{S. Tanaka, S. Ozaki, Y. Sakamoto, R. Tanuma, T. Yoshida and J. Murata}
%%\email[]{jiro@rikkyo.ac.jp}}

\address{Department of Physics, Rikkyo University, 3-34-1 Nishi-ikebukuro, Tokyo 171-8501, Japan}

\begin{abstract}
A new type of a trigger-scintillation counter array designed for the MTV experiment at TRIUMF-ISAC has been developed, which uses aluminum-metallized film tape for wrapping to achieve the required assembling precision of $\pm$0.5 mm. The MTV experiment uses a cylindrical drift chamber (CDC) as the main electron-tracking detector. The barrel-type trigger counter is placed inside the CDC to generate a trigger signal using 1 mm thick, 300 mm long thin plastic scintillation counters. Detection efficiency and light attenuation compared with conventional wrapping materials are studied. 
\end{abstract}

\begin{keyword} Plastic Scintillation Counter \sep Wrapping Material \sep Aluminum-Metallized Film
%% keywords here, in the form: keyword \sep keyword

%% MSC codes here, in the form: \MSC code \sep code
%% or \MSC[2008] code \sep code (2000 is the default)

\end{keyword}

\end{frontmatter}

%%
%% Start line numbering here if you want
%%
% \linenumbers

%% main text
\section{Introduction}
\label{intro}
The MTV experiment \cite{INPC,HYP} aims to perform the finest precision test of time reversal symmetry in nuclear beta decay by means of searching non-zero $T$-violating transverse polarization of the electrons emitted from polarized Li-8 nuclei, which is produced at TRIUMF-ISAC \cite{Levy}. This quantity may originate from the existence of a $T$-violating triple vector correlation defined as $R$-correlation in beta decay rate function \citep{Jackson}.
The existence of the $R$-correlation can be explored as the electron's non-zero transverse polarization, which is perpendicular to the parent nuclear polarization direction \cite{PSI}.
The electron's transverse polarization is measured as a backward-scattering left-right asymmetry from a thin (100 $\mu$m) lead analyzer foil using the known analyzing power of the Mott scattering \citep{Mott}.
The polarized Li-8 beam with 80\% horizontally polarized 28 keV beam at $10^7$ pps is irradiated on a surface of 10 $\mu$m thick aluminum beam stopper foil for which a spin relaxation time of 2.3 s is achieved using a pair of permanent magnets sandwiching the stopper foil placed in the horizontal direction to produce a spin-holding magnetic field of approximately 300 G.
The stopper is placed inside a 1.5 mm thick vacuum tube made of fiber reinforced plastics (FRP), designed to reduce the total material amount and mean atomic number $Z$ to suppress the multiple scattering of the low-energy electrons emitted from the beta decay.

\begin{figure}[h]
 \begin{center}
  \includegraphics[width=80mm]{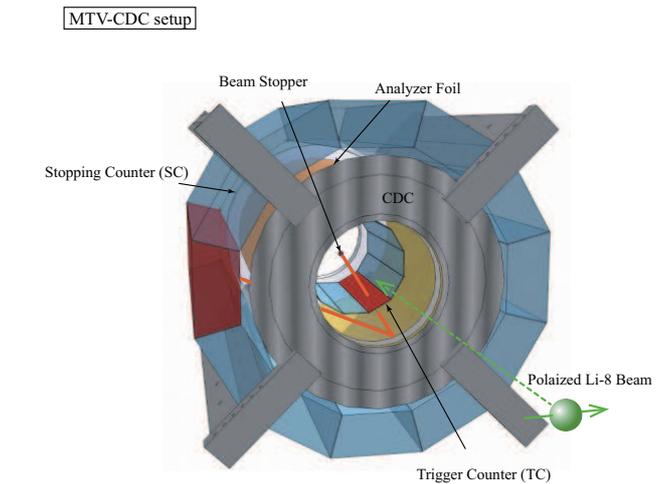}
 \end{center}
 \caption{Detector setup of the MTV experiment: trigger counter (TC), stopping counter (SC), and cylindrical drift chamber. The lead analyzer foil is placed around the CDC. The FRP beam stopper tube is not shown.}
 \label{MTV-setup}
\end{figure}

The electrons emitted from the Li-8 nuclei fly out from the FRP stopper tube, and their polarization is measured using the MTV detector array.
In addition to the Li-8 measurement, we also used this detector setup to perform a calibration measurement using an artificially produced transversely polarized electron source, and also, a gravity experiment named MTV-G \citep{MTV-G}, which probes large spin precession around the nuclei.
In the MTV-G and calibration measurements, the electron's polarization is measured in a double scattering experiment, using an un-polarized Sr (Y)-90 radiation source.
Longitudinally polarized electrons are emitted in nuclear beta decay, because of the parity violating weak interaction.
This longitudinal polarization is transferred to transverse polarization in Coulomb scattering at the first foil, and then, its transverse polarization is measured in the secondary Mott scattering analyzer foil \citep{MTV-G}.
The maximum beta energies of Li-8 and Y-90 are 13.1 MeV and 2.3 MeV, respectively, therefore, the absorbing, multiple scattering, and back-scattering effects of these low-energy electrons must be considered.

\begin{figure}[htbp]
 \begin{center}
  \includegraphics[width=70mm]{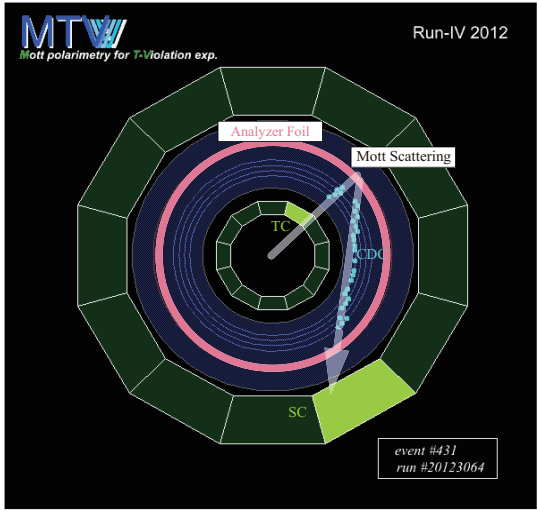}
 \end{center}
 \caption{A typical event display showing a V-track. The hit on the TC, scattering vertex position on the analyzer foil, and hits on the SC and CDC anode wires are shown.}
 \label{event-display}
\end{figure}

The MTV detector setup consists of the beam stopper, trigger counter (TC), lead analyzer foil, stopping counter (SC), and cylindrical drift chamber (CDC), as shown in Figure \ref{MTV-setup}.
The backward-scattering angular distribution is the measurement of transverse polarization, therefore, the CDC \citep{HYP} is set as the tracking chamber surrounding the stopper, which measures the incident tracks originating from the beam stopper and the scattered tracks starting at the point on the analyzer foil placed outside the CDC as V-shaped ("V-track") 
events. A typical V-track event is shown in Figure \ref{event-display}.Then electron's transverse polarization is obtained from the left and right scattering asymmetry.

\begin{figure}[htbp]
 \begin{center}
  \includegraphics[width=90mm]{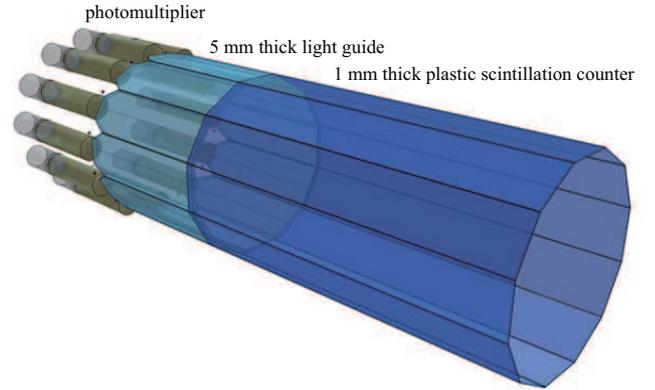}
 \end{center}
 \caption{Overview of the TC, which consists of 12 identical scintillation counter bars combined in a barrel configuration.}
 \label{TC-setup}
\end{figure}

The TC, which consists of 12 segmented 1 mm thick plastic scintillation counter bars (Figure \ref{TC-setup}), is placed inside the CDC, surrounding the FRP stopper vacuum tube.
The SC, which consists of 12 segmented 70 mm thick plastic scintillation counter bars, is placed outside of the CDC, which is designed to stop and measure the total kinetic energy of the scattered electrons.
The lead analyzer foil is placed between the CDC and SC in a cylindrical configuration, and its location is shown in Figure \ref{event-display}.
Therefore, an electron emitted from the beam stopper goes out into the air, penetrates one of the TC and CDC, and is then, backwardly scattered at the analyzer foil, from which it goes back to the CDC again. 
Finally, the electron stops at one of the SC bars after penetrating the analyzer foil at a position different from the scattered position.

The event trigger signal consists of two levels of triggering logics. 
The Level-1 trigger is generated with coincident signals between the signals of a TC and SC with the time window of 25 ns, while the Level-2 trigger requires multiple hits in the CDC anode wires after the Level-1 trigger.
The 12 segmentations of the TC and SC enable a rough selection of the backward-scattering event in the Level-1 triggering.

The prime requirements for the TC are supplying the input signal for the Level-1 trigger and working as a hodoscope array to select the rough direction of emission.
In addition, the TC works as a beam intensity monitor by providing a counting rate as well as an on-target beam polarimeter.
The transverse beam polarization can be determined by measuring the counting asymmetry between an pair of the TC counters set in an opposite configuration, utilizing parity-violating emission angular distribution.

\section{Trigger Counter (TC) Design}
\label{design}

The TC needs to satisfy the following requirements:
1) it must be as thin as possible to reduce the multiple scattering of the low-energy electrons around 1 MeV; 
2) it must be fast enough to generate the Level-1 input signal compared with the CDC's time scale of 100 ns; 
3) it must have a fine positioning to reduce the systematic effects on the $R$-correlation and MTV-G measurement.

\begin{figure}[htbp]
 \begin{center}
  \includegraphics[width=80mm]{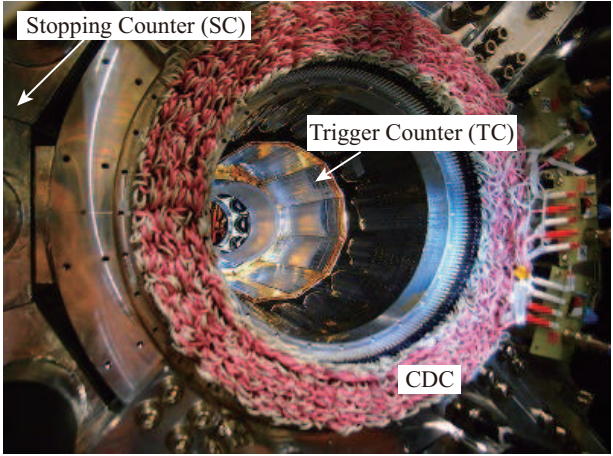}
 \end{center}
 \caption{Photograph of the TC installed inside the CDC. The FRP beam stopper tube is not shown.}
 \label{TC-setup-photo}
\end{figure}

To meet these requirements, we built 1 mm thick, 300 mm long thin plastic scintillation counter bars and assembled 12 such bars to form the TC barrel.
Each scintillation bar has 5 mm thick, 150 mm long acrylic light guide to transport the light output to the photomultipliers (PMTs) placed outside the CDC.
The details of the components of the TC are listed in Table \ref{scinti-list}.
The photograph of the TC is shown in Figure \ref{TC-setup-photo}.
PMT readout at one end is sufficient for the present experiment,
since timing resolution below coincidence window width is not required.
It is because timing selection is required only for reducing the  Level-1 trigger rate.
In addition, we do not read drift time information from the CDC, therefore, the TC dose not requested to provide starting signal with good timing resolution to achieve good position resolution from the drift chamber, which is dominated by multiple scattering.
However, the TC is designed to be possible to set PMTs on both ends as an upgrade plan.

\begin{table}[htbp]
\begin{center}
\begin{tabular}{|c||c|c|}
\hline
& material & size (mm) \\
\hline
\hline 
Scint. counter  & BC-408 & T.1 $\times$ W.50  $\times$ L.300  \\ 
\hline 
Light guide & acrylic & T.5 $\times$ W.50 $\times$ L.150\\ 
\hline 
PMT & Hamamatsu H-7415 & Dia.33 $\times$ L.130 \\
\hline
\end{tabular} 
\caption{Description of the TC scintillation counter bar, consisting of scintillation counter, light guide, and photomultiplier (PMT). (T.:thickness, W.:width, L.:length, and Dia.:diameter)}
\label{scinti-list}
\end{center}
\end{table}

\begin{figure}[htbp]
 \begin{center}
  \includegraphics[width=90mm]{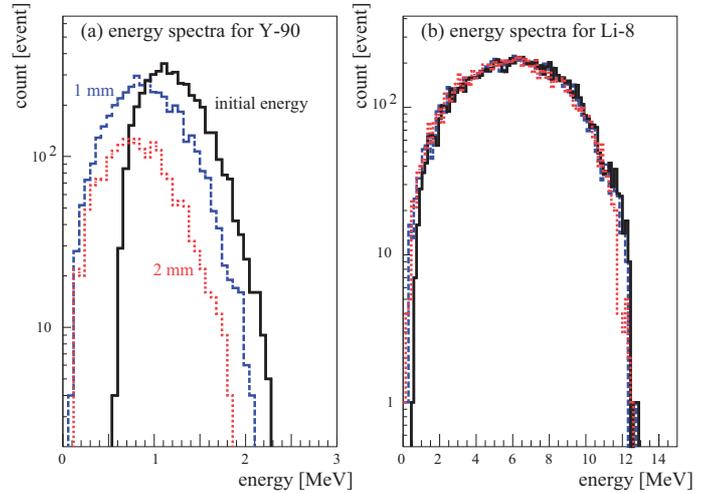}
 \end{center}
 \caption{Monte Carlo simulation of the energy of outgoing electrons through 1 mm and 2 mm thick scintillation counters for Y-90 and Li-8 beta decays. The electron's initial energy is also shown for comparison. Same numbers of events are generated for each case.}
 \label{Energy-MC}
\end{figure}

\begin{figure}[htbp]
 \begin{center}
  \includegraphics[width=90mm]{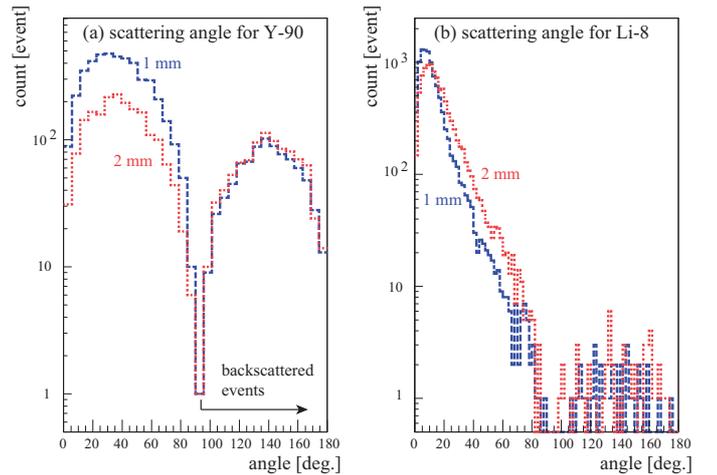}
 \end{center}
 \caption{Monte Carlo simulation of the scattering angular distributions of the outgoing electrons (angle $<90$ deg.), and of the backscattered electrons (angle $>90$ deg.) for Y-90 and Li-8 beta decays.}
 \label{Angle-MC}
\end{figure}

Before fixing the TC's design, in order to optimize the thickness of the TC, we performed a Monte Carlo simulation to evaluate the energy loss and multiple scattering effects from the scintillation counters (without wrapping materials) for the electrons emitted from the Li-8 and Y-90 beta decays.
In this custom-made Monte Carlo code, energy loss and multiple scattering effects are included.
Since the energy of electrons emitted from Sr-90 (maximum beta energy = 0.55 MeV) is small, it is ignored in this estimation.
In Figures \ref{Energy-MC} and \ref{Angle-MC}, the results of the Monte Carlo simulation for the energy of outgoing electrons after energy loss and their scattered angular distributions are shown, respectively.
Only those electrons whose energy was larger than 0.5 MeV were considered.
The number of penetrating electrons that were not stopped or backscattered at the scintillation counters was estimated.
The penetrating rates are listed in Table \ref{MC-result}.
Note that only 35\% of the incident electrons can penetrate the 2 mm plastic for Y-90.
The mean energy loss and scattering angles of forward scattering electrons were calculated and shown in Table \ref{MC-result}.
In the Y-90 measurements,
we found slight difference in the multiple scattering effects between the 1 mm and 2 mm thick samples, however, a large difference is expected on the stopping rate.
Therefore, the TC should be as thin as possible (approximately 1 mm) to maintain a good penetrating rate for the Y-90 measurements.
Considering the detection efficiency and machine ability, at least 1 mm thickness is required.
From this estimation, we decided to use 1 mm scintillation counters.
In case of Sr (Y)-90 measurements, we do not use the 1.5 mm thick FRP vacuum tube to reduce the material amount.
The Sr (Y)-90 source is set in air.

\begin{table}[h]
\begin{center}
\begin{tabular}{|c||c|c|}
\hline 
  & Y-90 & Li-8 \\ 
\hline 
\hline
penetrating rate for 1-mm & 86\% & 99\% \\ 
\hline 
penetrating rate for 2-mm & 35\% &  98\%\\
\hline 
mean scattering angle for 1-mm & 37 deg. & 12 deg. \\ 
\hline 
mean scattering angle for 2-mm & 38 deg. &  17 deg.\\
\hline
mean energy loss for 1-mm & 0.30 MeV & 0.19 Mev \\ 
\hline 
mean energy loss for 2-mm & 0.38 MeV &  0.31 MeV\\
\hline
\end{tabular} 
\caption{Monte Carlo estimation of the electron's penetrating rate,  mean scattering angle and mean energy loss for 1 mm and 2 mm thick plastic scintillation counters for Y-90 and Li-8 beta decays.}
\label{MC-result}
\end{center}
\end{table}

A key concern on using such thin and long plastic scintillation counters is their expected low efficiency due to light attenuation and small initial light yield.
On the other hand, the requirement of fine positioning has a drawback when we try to rely on total reflection by producing an air-layer surrounding the scintillation bar, which is required to make large refractive index difference at the surface.

\begin{figure}[htbp]
 \begin{center}
  \includegraphics[width=70mm]{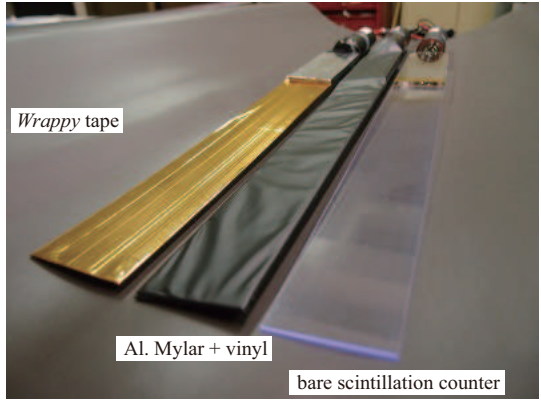}
 \end{center}
 \caption{Photograph of the TC, wrapped with
the {\it Wrappy} tape, the conventional aluminized Mylar with black vinyl sheet, and the bare scintillation bar. }
 \label{TCphoto}
\end{figure}

Therefore, we tried to use a new wrapping material made of aluminum-metallized tape, which has a good mirror-like reflecting surface on both sides of the tape.
We use “{\it Wrappy}” (CEMEDINE Co., Ltd., \cite{cemedine}) as the wrapping tape because of its mirror-like glossy back side using excellent transparent adhesive, which is also easily removable.
{\it Wrappy} is an aluminum-metallized polyester film tape, which consists of a 25 $\mu$m thick polyester film and good transparent acrylic adhesive.
The Total thickness of the {\it Wrappy} tape is $55\pm25\mu$m.
The {\it Wrappy} tape was directly glued on the surface of the plastic scintillation counters, as shown in Figure \ref{TCphoto}, where it is compared with a conventional aluminized Mylar wrapping with black vinyl sheet.
The geometrical precision after using the {\it Wrappy} tape in overlapping-wrapping is about $\pm$0.1 mm, which is sufficient for our application.

\section{Test measurement}
\label{test}

The main concern in using a direct taping on the surface of the scintillation counters without producing an air volume around the surface, is the expected reduction of light transmission to the PMTs.
For our usage, we needed to rely not on the total reflection at the index boundary between plastic and air, but on the reflection at the tape's evaporated surface.
Since the light-shielding ability of {\it Wrappy} is not sufficient for our requirements even with multiple overlapping, a 70 $\mu$m thick aluminum tape is added as a cover so that the remaining light leakage is negligible.
Direct taping of the aluminum tape without {\it Wrappy} is not good because its opaque glue with poor reflection, which is hard to remove from the plastic surface.

\begin{figure}[htbp]
 \begin{center}
  \includegraphics[width=80mm]{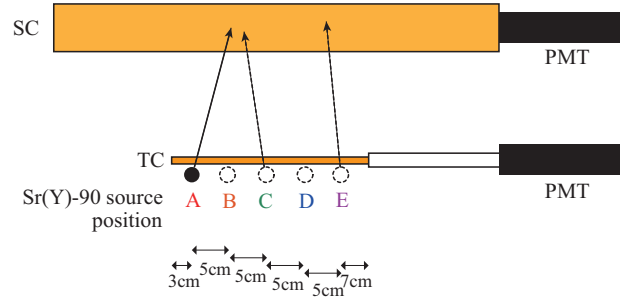}
 \end{center}
 \caption{Setup of the test measurement for studies of the study of the source position dependence. The QDC data from the TC and SC are shown.}
 \label{test-setup}
\end{figure}

\begin{figure}[htbp]
 \begin{center}
  \includegraphics[width=90mm]{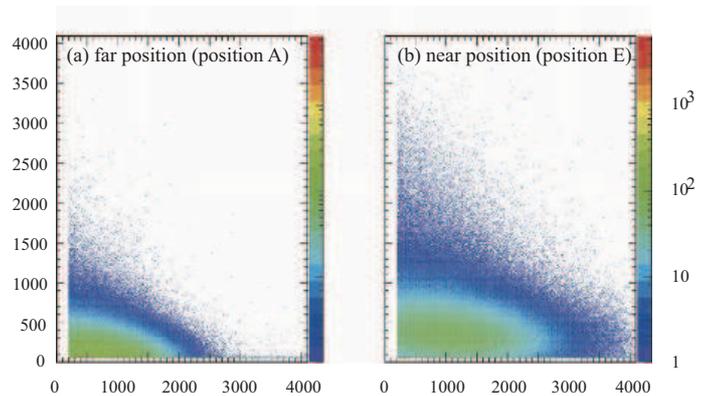}
 \end{center}
 \caption{PMT's pulse height distributions for the TC (vertical) and SC (horizontal) at (a) far and (b) near positions.}
 \label{SCTC-2d}
\end{figure}

Light attenuation was studied for the 1 mm thick TC.
In this test study, the output of the TC was measured in a test apparatus using an SC bar to produce trigger signals, as shown in Figure \ref{test-setup}.
The Sr (Y)-90 radiation source (3.7MBq) is attached on the TC with a collimator at five different positions (positions A to E, 5 cm spacing) to measure the transmitting length dependence on the output pulse height of the TC signal.
The output signals of the PMTs from the TC and SC were measured using a VME-QDC (CAEN V792), with the QDC gate generated by the SC placed behind the TC.
Figure \ref{SCTC-2d} shows the pulse height distributions from the TC and SC positioned at a long distance (position A) and short distance (position E).
The difference between them are considered as the result of the light attenuation in the TC.

\begin{figure}[htbp]
 \begin{center}
  \includegraphics[width=90mm]{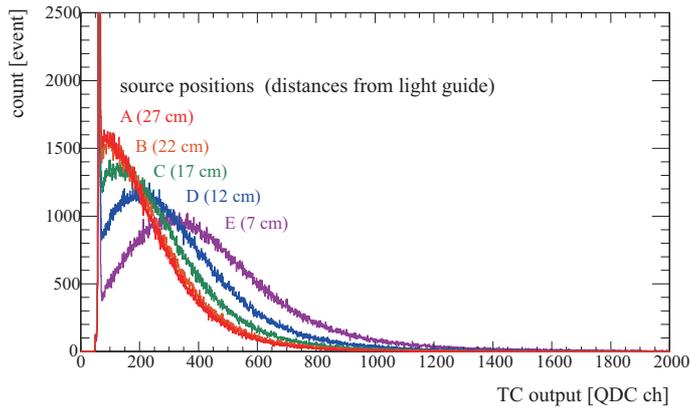}
 \end{center}
 \caption{PMT's output charge distributions of the TC at different source positions from A to E. Since the QDC gate signal is generated using the SC, this signal represents $\Delta E$ in the TC.}
 \label{1d-QDC}
\end{figure}

\begin{figure}[htbp]
 \begin{center}
  \includegraphics[width=90mm]{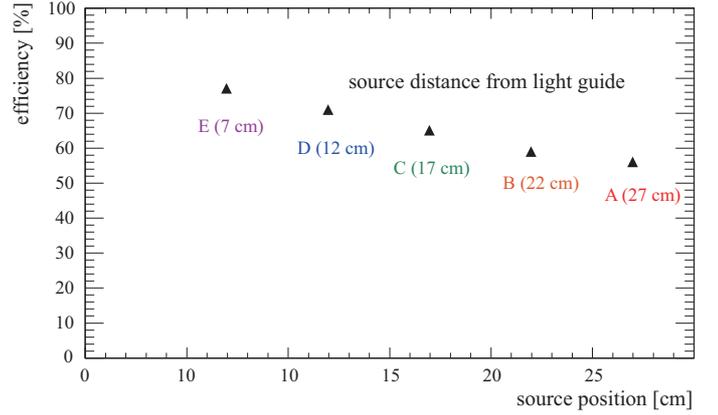}
 \end{center}
 \caption{Dependence of the source position on  detection efficiency of the TC.}
 \label{efficiency}
\end{figure}

The light attenuation behaviour can be clearly observed in one dimensional histograms shown in Figure \ref{1d-QDC}, where the pulse height distributions are shown for five different source positions.
As expected, a clear attenuation is observed, however, the attenuation of the pulse height itself is not an issue for the present application,
because the TC is not required to provide energy information, but is expected to generate the trigger signal.
Therefore, it is more important to investigate the detection efficiency.
Figure \ref{efficiency} shows the position dependence of the measured efficiency, defined as ($N_{TC \times SC})/N_{SC}$ after subtracting backgrounds.
The obtained result shows a relatively small efficiency reduction compared to the light attenuation shown in Figure \ref{1d-QDC}, indicating that the TC signal remains  significant above the noise level.
Although the obtained efficiency is not close to 100\%, it is a problem neither for the MTV nor MTV-G measurements,
because these experiments do not measure an absolute event rate, but measure the asymmetry during beam polarization direction flipping.
Therefore, in the double ratio analysis, most of the systematic effects from the efficiency difference are cancelled.
However, time variation of the efficiency causes systematic effects on the physics measurements, which should be carefully investigated.
Also, anisotropic efficiency distribution in azimuthal direction should be minimized to reduce the systematic error.

It should be remarked that, although a direct taping was used on the surface of the scintillation counters without producing an air-layer, no severe efficiency drop was observed in this measurement. 
As a result, we concluded that the thin 1 mm thick and 300 mm long plastic scintillation counters directly wrapped with {\it Wrappy} have sufficient efficiency for our application.

\section{Discussion}
\label{discussion}

\begin{figure}[htbp]
 \begin{center}
  \includegraphics[width=80mm]{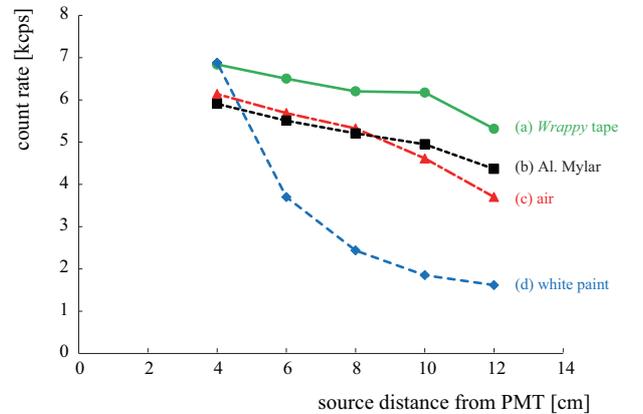}
 \end{center}
 \caption{Wrapping material effect on the counting rate measured for a 1 mm thick, 25 mm wide, and 140 mm long plastic scintillation counter for (a) {\it Wrappy} tape with aluminum tape, (b) aluminized Mylar with black vinyl sheet, (c) air (black vinyl sheet without aluminized Mylar), and (d) white paint. }
 \label{wrap-dep}
\end{figure}

The TC built for the MTV experiment proved to have sufficient absolute detection efficiency.
Independently, from the application for the MTV experiment,
we also studied the dependence of the wrapping material on the PMT's output in a dedicated measurement.
A plastic scintillation counter (1 mm in thickness, 25 mm in width, and 140 mm in length) was used to study this wrapping material dependence.
We tested (a) {\it Wrappy} with a 70 $\mu$m covering aluminum tape, (b) conventional aluminized Mylar with covering black vinyl sheet, (c) black vinyl sheet, and (d) white acrylic paint.
Figure \ref{wrap-dep} shows the counting rate measured in a setup similar to that shown in Figure \ref{test-setup}.
In this measurement, the source position dependence has been studied on five different source positions at a distance of 2 cm from each other.
It was found that {\it Wrappy} (a) was as good as the conventional aluminized Mylar wrapping (b).
Moreover, black vinyl sheet without aluminized Mylar (c) showed almost the same properties as that of the conventional material (b). 
It suggests that the total reflection between plastic and air is the dominant contribution to the light transmission.
Among all of them, the white painting on the plastic surface shows the strongest attenuation.

\begin{figure}[htbp]
 \begin{center}
  \includegraphics[width=90mm]{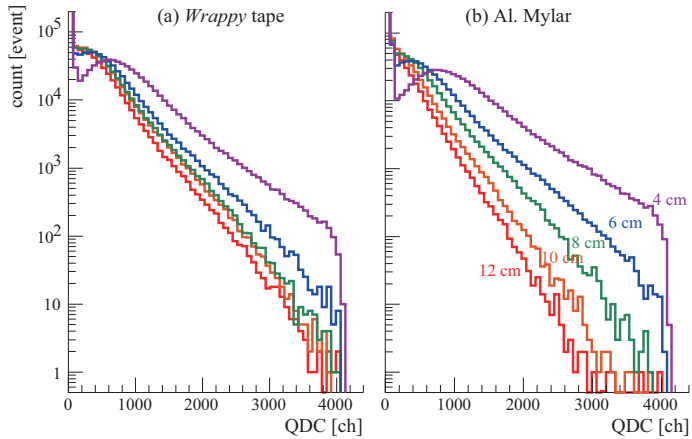}
 \end{center}
 \caption{Wrapping material effect on the light attenuation. QDC data are shown for {\it Wrappy} and aluminized Mylar cases, for different source positions of 4, 6, 8, 10, and 12 cm from the PMT. }
 \label{wrap-spectrum}
\end{figure}

The results of counting rates shown in Figure \ref{wrap-dep} depend on threshold settings, therefore, QDC spectra are better to be used for the comparison.
Indeed, the attenuation difference is more clearly seen in Figure \ref{wrap-spectrum}.
The QDC spectra of the materials containing (a) {\it Wrappy} and (b) aluminized Mylar at different source positions from the PMT are shown.
Aluminized Mylar showed a light attenuation stronger than of {\it Wrappy}.
Therefore, we can conclude that the {\it Wrappy} taping has excellent performance compared with the conventional wrapping materials.

\section{Conclusion}
\label{conclusion}
In summary, we attempted the direct taping on the surface of a plastic scintillation counter to improve the positioning precision for the MTV and MTV-G experiments.
Although, as expected, a slight attenuation was observed, the efficiency did not drastically drop, which would have negatively influenced the results of the experiments.
We found a new, easy, and reliable method to wrap scintillation counters, which can be used for many other experiments and applications.
For example, this new wrapping technique could be used for building fine segmented scintillation counter arrays, or scintillation counters placed inside vacuum chambers.

\section{Acknowledgements}
\label{ack}
This work was supported by TRIUMF's staffs under many technical aspects.
This work was supported by the JSPS KAKENHI Grant-in-Aid for Scientific Research (B) 25287061, Challenging Exploratory Research 24654070, and Rikkyo University Special Fund for Research (for Graduate Students 2013).

%% The Appendices part is started with the command \appendix;
%% appendix sections are then done as normal sections
%% \appendix

%% \section{}
%% \label{}

%% References
%%
%% Following citation commands can be used in the body text:
%% Usage of \cite is as follows:
%%   \cite{key}          ==>>  [#]
%%   \cite[chap. 2]{key} ==>>  [#, chap. 2]
%%   \citet{key}         ==>>  Author [#]

%% References with bibTeX database:

\bibliographystyle{model1-num-names}
\bibliography{<your-bib-database>}

%% Authors are advised to submit their bibtex database files. They are
%% requested to list a bibtex style file in the manuscript if they do
%% not want to use model1-num-names.bst.

%% References without bibTeX database:

\end{document}